\documentclass[conference,final]{IEEEtran}

\IEEEoverridecommandlockouts

\usepackage[pdftex]{graphicx}
\usepackage{amsmath,amssymb,amsfonts}
\usepackage{multirow,multirow}
\usepackage{makecell}
\usepackage{subcaption}
\usepackage[labelfont=bf,textfont=it,belowskip=0pt,aboveskip=5pt,tableposition=top]{caption}
\usepackage[table]{xcolor}
\usepackage[colorlinks=true,
citecolor=orange,
filecolor=black,
linkcolor=orange,
urlcolor=blue,backref=page]{hyperref}
\usepackage{algorithm,algorithmic}
\usepackage{siunitx}
\usepackage{booktabs}
\usepackage{times}
\newcounter{runidnum}
\newcommand{\runid}{\stepcounter{runidnum}\#\therunidnum}

\newcolumntype{R}{>{\columncolor{gray!20}}r}
\newcolumntype{L}{>{\columncolor{gray!20}}l}
\newcolumntype{C}{>{\columncolor{gray!20}}c}

\graphicspath{{figs}}
\DeclareGraphicsExtensions{.pdf,.png}

\newcommand{\figref}[1]{Figure~\ref{#1}}
\newcommand{\tabref}[1]{Table~\ref{#1}}
\newcommand{\secref}[1]{\S\ref{#1}}
\newcommand{\algref}[1]{Algorithm~\ref{#1}}

\newcommand{\vect}[1]{\boldsymbol{#1}}
\newcommand{\mat}[1]{\boldsymbol{#1}}
\newcommand{\ipoint}[1]{\textit{\textbf{#1}}}

\newcommand{\p} {\partial}
\newcommand{\Div}{\mbox{div}\,}
\newcommand{\Grad}{{\nabla}}
\newcommand{\Lap}{\rotatebox[origin=c]{180}{$\nabla$}}

\newcommand{\half}[1]{\frac{#1}{2}}

\newcommand{\MA}[1]{{\mathcal #1}}
\DeclareMathOperator{\bigO}{\mathcal{O}}

\definecolor{light-gray}{gray}{0.80}

\newcommand{\st}{\rho}          
\newcommand{\ad}{\lambda}       
\newcommand{\he}{\mat{\MA{H}}}  
\newcommand{\gr}{{\vect{g}}}    
\newcommand{\ve}{{\vect{v}}}    
\newcommand{\df}{{\vect{y}}}    
\newcommand{\po}{{\vect{x}}}    
\newcommand{\Po}{{\vect{X}}}    
\newcommand{\dt}{\delta t\,}    
\newcommand{\lea}{\mat{\MA{P}}} 
\newcommand{\vla}{\mat{\Lap}}   
\newcommand{\vbi}{\mat{\Lap^2}} 
\newcommand{\ivbi}{\mat{\Lap^{-2}}} 

\newcommand{\poi}{\po_{\vect{i}}} 
\newcommand{\Poi}{\Po_{\vect{i}}} 
\newcommand{\ipr}{\vect{i^\prime}}
\newcommand{\Pop}{\Po_{\ipr}} 

\newcommand{\ban}{t_w} 
\newcommand{\lat}{t_s} 

\newcommand{\z}{\phantom{0}}

\title{Distributed-Memory Large Deformation Diffeomorphic 3D Image Registration}
\author{
\IEEEauthorblockN{Andreas Mang, Amir Gholami, and George Biros}
\IEEEauthorblockA{
The Institute of Computational Engineering and Sciences\\
The University of Texas at Austin, Austin, Texas 78712--1229\\
andreas@ices.utexas.edu; amir@accfft.org; gbiros@acm.org
}

\thanks{This material is based upon work supported by AFOSR grants FA9550-12-10484 and FA9550-11-10339; by NSF grant CCF-1337393; by the U.S. Department of Energy, Office of Science, Office of Advanced Scientific Computing Research, Applied Mathematics program under Award Numbers DE-SC0010518 and DE-SC0009286; by NIH grant 10042242; by DARPA grant W911NF-115-2-0121; and by the Technische Universit\"{a}t M\"{u}nchen, Institute for Advanced Study, funded by the German Excellence Initiative (and the European Union Seventh Framework Programme under grant agreement 291763). Any opinions, findings, and conclusions or recommendations expressed herein are those of the authors and do not necessarily reflect the views of the AFOSR, the DOE, the NIH, the DARPA, or the NSF. Computing time on the Texas Advanced Computing Centers Stampede system was provided by an allocation from TACC and the NSF.}
}

\IEEEoverridecommandlockouts
\IEEEpubid{
\parbox[t]{\columnwidth}{
accepted for publication at SC16; Salt Lake City, Utah, USA; November 2016\\
\hfill} \hspace{\columnsep}\makebox[\columnwidth]{}
}

\begin{document}
\bstctlcite{IEEEexample:BSTcontrol}
\maketitle

\begin{abstract}
We present a parallel distributed-memory algorithm for large deformation diffeomorphic registration of volumetric images that produces large isochoric deformations (locally volume preserving). Image registration is a key technology in medical image analysis. Our algorithm uses a partial differential equation constrained optimal control formulation. Finding the optimal deformation map requires the solution of a highly nonlinear problem that involves pseudo-differential operators, biharmonic operators, and pure advection operators both forward and backward in time. A key issue is the time to solution, which poses the demand for efficient optimization methods as well as an effective utilization of high performance computing resources. To address this problem we use a preconditioned, inexact, Gauss-Newton-Krylov solver. Our algorithm integrates several components: a spectral discretization in space, a semi-Lagrangian formulation in time, analytic adjoints, different regularization functionals (including volume-preserving ones), a spectral preconditioner, a highly optimized distributed Fast Fourier Transform, and a cubic interpolation scheme for the semi-Lagrangian time-stepping. We demonstrate the scalability of our algorithm on images with resolution of up to $1024^3$ on the ``Maverick'' and ``Stampede'' systems at the Texas Advanced Computing Center (TACC). The critical problem in the medical imaging application domain is strong scaling, that is, solving registration problems of a moderate size of $256^3$---a typical resolution for medical images. We are able to solve the registration problem for images of this size in less than five seconds on 64 x86 nodes of TACC's `'Maverick'' system.
 \end{abstract}


\section{Introduction}\label{s:intro}

\begin{figure*}
\includegraphics[width=.99\textwidth]
{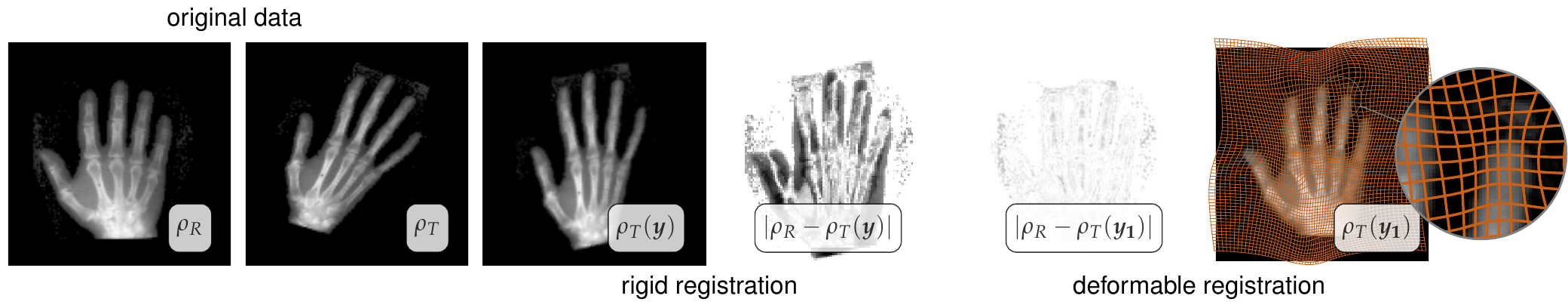}
\caption{The image registration problem (data taken from \cite{Modersitzki:2009a,Amit:1994a}). The input (original data) are image intensities $\st_R$ and $\st_T$. The output is $\df$, the deformation map. Our goal is to find $\df$ so that $\st_T(\df)$ (the \ipoint{deformed} $\st_T$) is as close as possible to $\st_R$ (with respect to some appropriate measure). One way to achieve this is to use rigid registration (i.e., searching for a map that entirely is described by rotations and translations). The result of a rigid registration is shown in the third image from the left; the fourth image shows the difference between the two images $\st_R$ and $\st_T$ after rigid registration. As one can see, there are still significant differences in the intensity values. If we use a deformable registration instead, we can compute a much more flexible $\df_1$, which results in a much smaller misfit $|\st_R-\st_T(\df_1)|$. The deformation map $\df_1$ for this method is visualized in the last figure to the right. The grid lines, superimposed on top of $\st_T(\df_1)$ were a Cartesian grid before the deformation. We use them to visualize the overall deformation.}
\label{f:regist101}
\end{figure*}

\begin{figure}
\includegraphics[width=0.99\linewidth]
{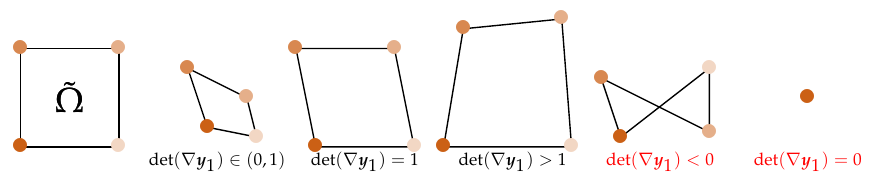}
\caption{Here we illustrate diffeomorphic and non-diffeomorphic transformations in 2D by considering an infinitesimal area $\tilde{\Omega}$. We can think of $\tilde{\Omega}$ as a single grid cell (pixel). The first figure on the left depicts the undeformed area. The second figure shows an admissible deformation that shrinks the original area. The third figure depicts an area-preserving deformation (in 3D it would be volume preserving). The fourth figure shows the result of a deformation that expands the area. The fifth illustration is not diffeomorphic since material lines that did not cross each other in the original (leftmost illustration), cross now. The sixth, rightmost, is illustration corresponds to a singular deformation in which all the spatial information is lost by shrinking $\tilde{\Omega}$ to a single point. The last two deformations are not useful in image analysis. However, without any constraints on the deformation map, pixels with non-diffeomorphic behavior appear almost always. For this reason appropriate regularization of the problem is necessary.}
\label{f:diffeo}
\end{figure}

Deformable registration (also known as image alignment, warping, or matching) refers to methods that find point correspondences between images by comparing image intensities. We refer to these point correspondences as the \ipoint{deformation map} (see \figref{f:regist101} in \secref{s:background}). An example for a low dimensional image registration problem is affine registration; it creates simple maps consisting of rotations, translations, and scalings~\cite{Modersitzki:2004a}. Typically, affine registration is used as an initialization step for \ipoint{large deformation diffeomorphic registration} ({\bf LDDR}), which is the problem we are concerned with in the present work. In LDDR, we typically search for a deformation map for which the degrees of freedom are the ambient space times the number of grid points defined in the image space. LDDR is much more flexible than affine registration and thus, in general, more informative in clinical studies~\cite{Sotiras:2013a,Klein:2009a}. Such high-dimensional transformations can be defined in many different ways~\cite{Modersitzki:2004a,Modersitzki:2009a,Sotiras:2013a}. Image registration is an ill-posed inverse problem; it does not have a unique solution. Not all large deformation maps are \emph{admissible} since they can shuffle the points arbitrarily to match intensities. It is crucial to impose constraints on the deformation while allowing for flexibility. The most important constraint is that the map is \ipoint{diffeomorphic} (see~\figref{f:diffeo} in \secref{s:background}).

\IEEEpubidadjcol

Solving an LDDR problem in a rigorous way requires the solution of a non-convex partial differential equation ({\bf PDE}) constrained optimal control problem \cite{Borzi:2012a,Gunzburger:2003a,Hinze:2009a,Lions:1972a}. This problem is ill-posed and involves non-linear and ill-conditioned operators. Most state-of-the-art packages circumvent these issues by sacrificing scalability and settling for crude solutions using simple but suboptimal algorithms. In many cases this works sufficiently well, but in several other cases, it does not. There is significant activity in trying to improve the existing algorithms. With regards to performance optimizations, most codes use open multi-processing ({\bf OpenMP}) or graphics processing unit ({\bf GPU}) acceleration; there are very few codes that utilize distributed memory parallelism. As a result they are not scalable to the full resolution; to solve problems for large images most codes use subsampling. This is limiting considering the current imaging resolutions. Seven Tesla magnetic resonance imaging ({\bf MRI}) scanners can reach a resolution of up to \SI[round-precision=1]{0.5}{\milli\metre} ($\approx 450^3$ voxels)~\cite{Lusebrink:2013a}. Ultra-high resolution computed tomography ({\bf CT}) captures \SI[round-precision=2]{0.25}{\milli\metre} resolution ($\approx512^3$ voxels)~\cite{Kakinuma:2015a}.

Beyond the need for strong scaling of image registration algorithms for clinical applications, there is  also need for weak scaling for imaging in biology, biophysics, and neuroscience.
Animal Micro-CT reaches $\bigO(\si{\micro\metre})$ resolution ($\approx2000^3$ voxels)~\cite{Starosolski:2015a,Zhang:2015a}. In small animal neuroimaging, CLARITY~\cite{tomer-e14}, a novel optical imaging technique,  can deliver sub-micron resolution for the whole brain of the animal, resulting in 10-100 billion-voxel images.  To our knowledge, none of the existing schemes for LDDR allow for the registration of such large volumetric images~\cite{Ashburner:2011a,Beg:2005a}.

\ipoint{Contributions}: The design goals for our 3D LDDR scheme are the following: (1) ability to represent large diffeomorphic deformations; (2) algorithms based on rigorous mathematical foundations; (3) algorithmic optimality with respect to \emph{both} the deformation map resolution and the image resolution; and (4) parallel scalability. Here, we propose an algorithm that achieves these goals and has the following characteristics:

\begin{itemize}
\item It is based on optimal control theory. Our formulation allows the control of the registration quality in terms of image correspondence and different quality metrics for the diffeomorphism/deformation map (see~\secref{e:registration}).
\item It uses a semi-Lagrangian approach for solving the transport equations that govern the deformation of the image. This approach leads to algorithmic optimality (see~\secref{s:time}).
\item It uses a spectral discretization in space (see~\secref{s:space}). This discretization enables flexibility in the choice of regularization operators for the deformation map. Such flexibility is necessary since different image registration applications have different requirements. It also allows for efficient solvers of saddle-point linear systems.
\item It uses optimal algorithms based on an adjoint-based formulation solved via a line-search globalized, inexact, preconditioned Gauss-Newton-Krylov scheme (see~\secref{s:newton}).
\item It uses distributed-memory parallelism for scalability, employs several performance optimizations specific to our problem, and uses a parallel FFT for elliptic solvers and differentiations that has been shown to scale to hundreds of thousands of cores (see~\secref{s:hpc}). It employs several optimizations for the most expensive part of the computation (cubic interpolation) (see~\secref{s:interp}). It also supports GPU acceleration (not discussed here).
\end{itemize}

\noindent In addition, we analyze the overall complexity of our method in terms of communication, computation, and storage. The class of deformations we consider here are one of the most challenging since we enable locally volume preserving maps,\footnote{In the medical imaging jargon this is referred to as \emph{``mass preserving''} maps.} which find many applications~\cite{Shen:2003b,Yin:2009a,Haber:2004a,Burger:2013a}. We present results for synthetic and neurological images and demonstrate the performance of our algorithm (see~\secref{s:results}) for both volume preserving and more generic deformation maps.

\ipoint{Related work on high performance computing methods for 3D image registration}: A rich literature survey on high performance computing ({\bf HPC}) in image registration can be found in~\cite{Shams:2010a,Eklund:2013a,Shackleford:2013a}. General surveys on image registration can be found in~\cite{Modersitzki:2004a,Sotiras:2013a}. Formulations related to the one discussed in this work are reviewed in~\cite{Mang:2015a,Mang:2016a,Mang:2016b}.

State of the art registration packages that are used in the medical imaging and medical image computing community include ELASTIX~\cite{Klein:2010a}, ANTS~\cite{Avants:2011a}, DARTEL \cite{Ashburner:2007a}, and DEMONS~\cite{Vercauteren:2008a,Vercauteren:2009a,Lorenzi:2013b}. All of these offer some kind of diffeomorphic registration scheme. These packages mostly support OpenMP, but do not use GPUs or Message Passing Interface ({\bf MPI}) acceleration (exceptions to be discussed below). An important distinction should be made between the \emph{image resolution} (number of voxels) and the \emph{map resolution} (number of degrees of freedom for the map parameterization). In general, the higher the map resolution, the better the registration quality~\cite{Klein:2009a,Klein:2010a} but the harder the optimization problem since it has more degrees of freedom. Most existing codes downsample the map resolution significantly.

The majority of researchers have used GPUs to accelerate the calculations. For example, the solver for the LDDR scheme in~\cite{Ha:2010a} uses a preconditioned gradient descent (not Newton) algorithm with a hardware-provided trilinear interpolation on a GPU architecture. It supports the distributed solution of multiple independent LDDR image registration problems (in an embarrassingly parallel way), but does not support distributed memory parallelism for a single LDDR problem.

Two popular packages that exploit GPU acceleration are NIFTYREG~\cite{Modat:2010a} and PLASTIMATCH~\cite{Schackleford:2010a}. They use B-spline parameterized low-resolution maps ($50^3$ coefficients), a tri-linear interpolation scheme, and gradient descent type optimization; NIFTYREG supports soft constraints to penalize volume change.

An MPI version of NIFTYREG for bigger images~\cite{Ino:2005a} exists, but the map resolution remains the same ($50^3$ regular grid for the deformation field). The pioneering works~\cite{Warfield:2000a,Liu:2009a} support MPI. Their formulation is based on elastic deformation maps (not LDDR).  A GPU-LDDR scheme that supports somewhat high-resolution maps ($128^3$) is~\cite{Rehman:2009a}. It uses steepest descent (not Newton) and does not support MPI; no timings are reported.

To summarize, existing schemes do not support scalable LDDR algorithms and no scaling studies have been reported.

\ipoint{Limitations}: In multiframe volume registration (e.g., 4D Cine-MRI) one seeks to register multiple images using a smooth, continuous mapping~\cite{Chen:2011a,Beg:2005a}. Our solver can be used as is, but our diffeomorphic map parameterization is better suited for registering two images. Our parameterization can be extended without any major algorithmic changes but the software engineering would require some work. Another missing piece is a preconditioner that is insensitive to the regularization parameter. There are several techniques for doing so, e.g., grid continuation and multilevel preconditioning~\cite{Borzi:2012a,Adavani:2008b,Mang:2016b}. Here we focus on the single-level solver. The single node performance of the interpolation can be improved by more sophisticated blocking, manual vectorization, and possibly prefetching. Similar considerations hold true for the GPU version of the interpolation. This is ongoing work. Another limitation is that we only consider a discretization on Cartesian grids. This is not always the best grid~\cite{Warfield:2000a}. The structure of our algorithm changes significantly for unstructured grids.

\section{Background}\label{s:background}

Let $\Omega =[0,2\pi)^3\subset\mathbb{R}^3$ be the spatial domain, in which we define functions (images). $\p\Omega$ denotes the boundary of $\Omega$ and $\po$ a point in $\Omega$. Let $\st(\po)\in\mathbb{R}$ be a function defined on $\Omega$. In imaging $\st(\po)$ is the \ipoint{image intensity} at a point $\po$; in optimal control $\st(\po)$ is the \ipoint{state field}. In the registration problem, we have a reference image, denoted by $\st_R(\po)$, and a template image, denoted by $\st_T(\po)$; the goal is to find a vector valued deformation map, denoted by $\df(\po)$, that maps a point of the template image $\st_T$ to a corresponding point in the reference image $\st_R$~\cite{Modersitzki:2004a,Modersitzki:2009a}.

Let $\ve(\po)$ be the \ipoint{velocity field} that generates the map $\df$. In our formulation, we introduce a \emph{pseudotime} to denote the deformation of the template image at time $t$, denoted by $\st(\po,t)$. We define $\st(\po,t=0)$ to be the undeformed template image $\st_T$, and $\st(\po,t=1)$ to be the result of applying the deformation map (which needs to be compared to $\st_R(\po)$).

For the optimal control problem, $\lambda(\po,t)$ is the \ipoint{adjoint field}, $\he$ is the reduced \ipoint{Hessian operator}, $\gr$ is the \ipoint{gradient field}, and $\beta>0$ is the scalar regularization parameter. We use periodic boundary conditions for all differential operators. For the discretization, $N_i$ is the \ipoint{number of grid points} per $i$-th dimension; $N_1 N_2 N_3$ is the total number of unknowns in space; $n_t$ is the \ipoint{number of time steps} and $\dt$ the \ipoint{time step size}. We use $p$ for the number of \ipoint{MPI tasks}, $\lat$ for the latency in seconds, and $\ban$ for the reciprocal of the bandwidth when we do complexity analysis. Boldface lowercase symbols indicate vectors in $\mathbb{R}^3$.

\subsection{Image registration}\label{e:registration}

The image registration problem can be abstractly defined as follows. Given two functions (i.e., images) $\st_T(\po)$ and $\st_R(\po)$, we seek a vector function $\df_1(\po)$ such that the $L^2$-distance (i.e., the residual) between $\st_T(\df_1(\po))$ and $\st_R(\po)$ is minimal. We can think of $\df_1(\po)$ as deforming an (infinite) grid of points in  the template image $\st_T$ so that their intensity after the deformation matches the reference image $\st_R$ in the $L^2$-norm.\footnote{Other types of \emph{distance measures} can be used (see, e.g.,~\cite{Modersitzki:2004a,Modersitzki:2009a,Sotiras:2013a}). There are no significant changes in our formulation or algorithm if we would consider other, popular distance measures.}

It can be shown that this problem has an infinite number of solutions, most of which are not useful. To resolve this, one needs to impose additional constraints, such as smoothness (i.e., $\Grad\df_1(\po)$ exists), although this alone may not be sufficient to ensure that the map is plausible. Note, that we require $\det(\Grad\df_1(\po))>0$ for all $\po\in\Omega$ to guarantee that $\df_1(\po)$ is a \emph{diffeomorphism} (for an illustration, see~\ref{f:diffeo}). In velocity-based LDDR it can be shown that such a diffeomorphic map $\df_1$ exists, if the generating \ipoint{velocity field} $\ve(\po,t)$ is adequately smooth~\cite{Beg:2005a,Crippa:2007a,Chen:2011a}; a typical requirement is that $\ve$ is an $H^2$-function~\cite{Beg:2005a}. The deformation map $\df_1$ can be computed from $\ve$ by solving
\begin{equation}\label{e:euler-def}
\p_t\df(\po,t)+\ve(\po,t)\cdot\Grad\df(\po,t) = 0,\quad \df(\po,0)=\po,
\end{equation}

\noindent where $\df_1(\po)=\df(\po,1)$. It can be shown that if the velocity is \emph{incompressible} (i.e., $\Div \ve = 0$ for every $\vect{x}$ in $\Omega$), then $\det\Grad\df(\po,t)=1$ and the diffeomorphism is referred to as \emph{volume preserving}~\cite{Gurtin:1981a}. Here, we consider both the general and the incompressible velocity cases. The latter case is more challenging. We only consider stationary velocity fields, that is, $\ve(\po,t)=\ve(\po)$.\footnote{It can be shown that the space of possible diffeomorphisms generated by time-varying velocities is strictly larger than the space generated by stationary velocities. This does not have practical implications when we register two images (see, e.g., \cite{Mang:2015a}). However, it is restrictive if we want register  a sequence of images, like in optical flow problems.}

In the following we drop the dependence of the functions on the spatial position $\po$ for notational convenience.

\subsection{Formulation}\label{e:formulation}

The solution to the image registration problem can be found by solving the following PDE-constrained optimization problem~\cite{Hart:2009a,Chen:2011a,Mang:2015a}:
\begin{subequations}
\label{e:varopt}
\begin{equation}
\min_{\ve} \MA{J}[\ve]=
\half{1}\|\st_1 - \st_R\|^2_{L^2(\Omega)}
+ \half{\beta} \| \vla\ve\|^2_{L^2(\Omega)}
\end{equation}
\noindent subject to
\begin{align}
\p_t \st(t) + \ve\cdot\Grad \st(t)& = 0
&& {\rm in}\;\Omega\times(0,1], \label{e:varopt:state}
\\
\st(0) &= \st_T
&& {\rm in}\;\Omega,
\label{e:varopt:state-ic}
\\
\Div \ve
&=0
&& {\rm in} \; \Omega.
\label{e:varopt:constraint}
\end{align}
\end{subequations}

\noindent The second term of $\MA{J}$ enforces smoothness for $\ve$ and $\beta>0$ is the \emph{regularization parameter}. In this formulation, $\st_1(\po) = \st(\po,1)\equiv\st_T(\df_1)$, where $\df_1$ is the solution of~\eqref{e:euler-def} at $t=1$. The constraint~\eqref{e:varopt:state} defines an implicit function between $\st_1$ and $\ve$; given $\ve$ we solve~\eqref{e:varopt:state} for $\st$.

\paragraph{Computing the gradient of $\MA{J}$}

Given $\ve$, we need several steps to compute the gradient $\gr=\p_{\ve}\MA{J}$. First we compute $\st(1)$ by solving~\eqref{e:varopt:state} (with initial condition defined by \eqref{e:varopt:state-ic}). Then we compute the adjoint function $\ad(t)$ by solving the \ipoint{backward-in-time adjoint equation}~\cite{Ito:2008a,Borzi:2012a}
\begin{align}\label{e:adjoint}
-\p_t \ad(t) - \Div (\ve \ad(t)) & = 0
&& {\rm in}\;\Omega\times [0,1),
\\
\ad(1) = \st_R - \st(1)&
&& {\rm in}\;\Omega.
\nonumber
\end{align}

\noindent Once we have the state and adjoint fields, we can evaluate the \ipoint{gradient} given by
\begin{equation}\label{e:gradient}
\gr(\ve):= \beta \vbi \ve +
(I-\Grad\Lap^{-1}\Div)\int_0^1 \ad(t)\Grad\st(t)\,\mathrm{d}t.
\end{equation}

\noindent The operator $\lea=I-\Grad\Lap^{-1}\Div$, also known as the \emph{Leray operator}~\cite{Temam:1977a}, eliminates the incompressibility constraint for $\ve$. Furthermore, we define the vector field \[\vect{b}(\po) := \int_0^1\ad(\po,t)\Grad\st(\po,t)\,\mathrm{d}t,\] so that $\gr(\ve)=\beta\vbi\ve+\lea\vect{b}$. The gradient $\gr$ is a nonlinear elliptic integro-differential operator where the state (forward in time) and adjoint (backward in time) transport PDEs are ``hidden'' in $\vect{b}$.

Evaluating the gradient requires solving two transport equations, inverting the Laplacian and applying gradient, divergence, and biharmonic operators. (If we do not wish to compute a volume preserving map we can drop (i.e., not enforce) the incompressibility constraint. Then, in the gradient calculation, we only need to  replace the $\lea$ operator with an identity operator.) \ipoint{The first-order optimality condition} for~\eqref{e:varopt} requires that $\gr(\ve)=\vect{0}$. Most registration packages use steepest descent (first order) methods to find an optimal point (minimizer)~\cite{Avants:2011a,Beg:2005a,Chen:2011a}. However, steepest descent methods only have a linear convergence rate. Using Newton methods, which provide a much better convergence rate, is considered to be prohibitive, especially for LDDR for two main reasons. First, it is cumbersome to derive the equations for the second order optimality conditions. Second, a naive implementation of Newton methods can be very costly if not done carefully.

\paragraph{The Newton and Gauss-Newton Hessian operators $\he$}

To solve $\gr(\ve)=\vect{0}$ for $\ve$ we use an Armijo line-search globalized Newton method~\cite{Nocedal:2006a}. The key operation is the action of the \ipoint{Hessian} $\he(\ve)$ on a vector field $\tilde{\ve}$, which is commonly referred to as the \ipoint{Hessian matvec}. This matvec is computed by performing the following steps: First of all, we need to solve \eqref{e:varopt:state} and \eqref{e:adjoint} to compute the state and adjoint variables $\st$ and $\ad$, respectively. After computing these fields, we need to solve~\eqref{e:inc-state} for the \emph{incremental state} variable $\tilde{\st}$ and~\eqref{e:inc-adj} for the \emph{incremental adjoint} variable $\tilde{\ad}$, accumulating them in time to compute $\tilde{\vect{b}}$ and finally evaluating~\eqref{e:inc-control}.
\begin{subequations}
\label{e:hessian}
\begin{IEEEeqnarray}{rcl}
\IEEEyessubnumber*
\p_t \tilde{\st}(t) + \ve\cdot\Grad\tilde{\st}(t)
+ \tilde{\ve}\cdot \Grad \st(t) \; & = 0\
&\text{in}\;\; \Omega \times (0,1],
\label{e:inc-state}
\\
\tilde{\st}(0)\; &= 0\
&\text{in}\;\;\Omega,
\label{e:inc-state-ic}
\\
-\p_t \tilde{\ad}(t)
-\Div(\tilde{\ad}(t)\ve+\ad(t)\tilde{\ve})
\;& = 0\
&\text{in}\;\; \Omega \times [0,1),
\label{e:inc-adj}
\\
\tilde{\ad}(1) +\tilde{\st}(1)\; &= 0\
&\text{in}\;\;\Omega,
\label{e:inc-adj-fc}
\\
\he(\ve)\tilde{\ve}:=
\beta\vbi\tilde{\ve}
+ \lea\tilde{\vect{b}}
&&\text{in}\;\;\Omega,
\label{e:inc-control}
\end{IEEEeqnarray}
\end{subequations}
\noindent where
\[
\tilde{\vect{b}} =
\int_0^1 {\tilde{\ad}(t) \Grad \st(t)
+ \ad(t) \Grad \tilde{\st}(t)}\,\mathrm{d}t.
\]

Notice that \eqref{e:inc-state} and \eqref{e:inc-adj} require storing $\st(t),\tilde{\st}(t)$, and $\ad(t)$ for all $t$. Also notice that certain terms in \eqref{e:hessian} drop if we enforce $\Div\ve=0$. The \ipoint{Newton step}, $\tilde{\ve}$, is obtained by \ipoint{solving the linear system} $\he(\ve)\tilde{\ve}=-\gr(\ve)$. For a \ipoint{Gauss-Newton} Hessian, we drop the two terms that involve $\ad(t)$, i.e., the last term in \eqref{e:inc-adj} and the second term in $\tilde{\vect{b}}$.

\section{Methods}\label{s:methods}

\begin{figure}
\centering
\includegraphics[width=0.75\linewidth]
{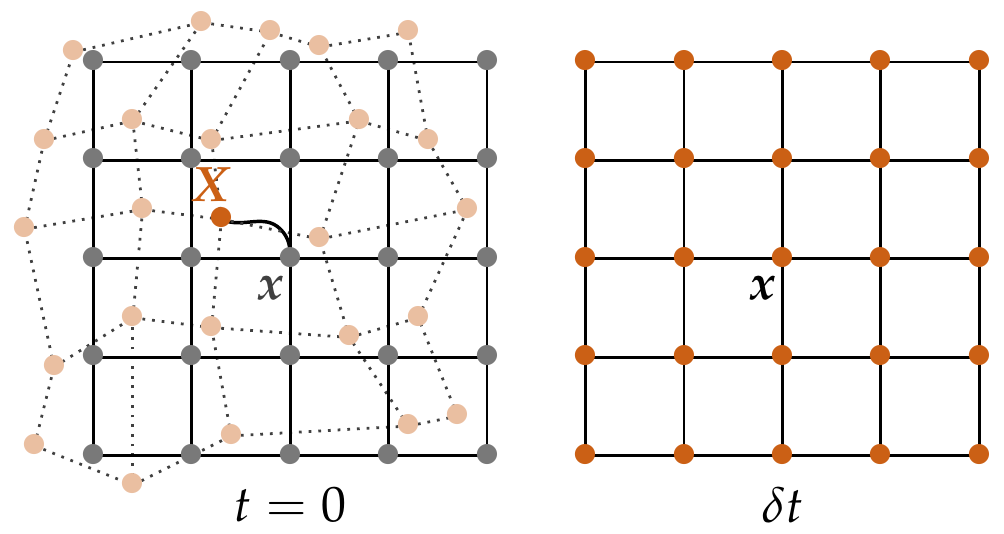}
\caption{Illustration of the semi-Lagrangian scheme (figure modified from \cite{Mang:2016b}). The square domain corresponds to a grid block assigned to a single MPI task. Point $\po$ is a regular grid point and $\Po$ is the material point at $t=0$ that landed at $\po$ at time $\dt$. The semi-Lagrangian algorithm requires interpolation of $\ve,\st,\tilde{\st},\ad$, and $\tilde{\ad}$ at these off-grid points; as illustrated here, these points can lie in other MPI tasks. We communicate these points, interpolate their values, and then communicate them back.}
\label{f:smlpoints} \end{figure}

Given two images $\st_R$ and $\st_T$ our goal is to find $\df_1$.
We use an optimize-then-discretize approach for \eqref{e:varopt}.
Our scheme can be summarized as follows:
\begin{itemize}
\item We solve \eqref{e:varopt} for $\ve$.
\item Once we have $\ve$, we use \eqref{e:euler-def} to compute $\df_1$.
\item To find $\ve$ we solve $\gr(\ve)=\vect{0}$ (where $\gr(\ve)$ is given by \eqref{e:gradient}) using a preconditioned Newton-Krylov method.
\item In space we use  Fourier expansions (regular grids with periodic boundary conditions) and in time we use a semi-Lagrangian scheme.
\item We use data parallelism in space (the regular grid).
\item All spatial differential operators (and their inverses if needed) are computed spectrally using our parallel FFT.
\item All spatial algebraic operators are done in parallel.
\end{itemize}

\noindent We provide more details for our algorithm in the following subsections.

\subsection{Newton-Krylov solver and preconditioning}\label{s:newton}

For the optimization we use a Newton method globalized with an Armijo line-search. We use a preconditioned Conjugate-Gradient (PCG) method to compute the Newton step. The linear solves using PCG are done inexactly using a tolerance that depends on the relative norm of the gradient~\cite{Eisenstat:1996a}. The preconditioner is the inverse of the biharmonic operator ($\ivbi$) and can be applied in nearly linear time using FFTs (with a logarithmic factor). This preconditioner delivers mesh-independence---but not $\beta$-independence (see \secref{s:results}). Since the problem is highly nonlinear we use parameter continuation on $\beta$. The target value for $\beta$ is application dependent and, in our algorithm, determined by various metrics defined on $\Grad\df_1$~\cite{Mang:2015a}. We use the {\tt TAO} module from the {\tt PETSc} library~\cite{Balay:2016a,petsc-web-page} for numerical optimization, which supports user-defined, matrix free PCG. {\tt TAO} provides interfaces that allow one to control two main parameters in the Newton-Krylov solver: ($i$) the accuracy of the solution of the linear system to compute the Newton step (the relative tolerance of the PCG method used to solve the Hessian equation); and ($ii$) the nonlinear termination criteria. We provide the algorithms to determine these parameters and, given $\ve$ and $\tilde{\ve}$, efficient routines for the function evaluation $\MA{J}(\ve)$, the gradient evaluation $\gr(\ve)$, the Hessian matvec $\he({\ve})\tilde{\ve}$ and the action of the preconditioner $\ivbi\tilde{\ve}$.

\subsection{Discretization}\label{s:discretization}

We use Cartesian grids to approximate spatial functions and spatial differential operators. We use an explicit Runge-Kutta second-order semi-Lagrangian method to discretize in time.

\subsubsection{Space}\label{s:space}

We use a spectral projection scheme for all spatial operations on a regular grid defined on $\Omega$ with periodic boundary conditions. For simplicity, we consider the isotropic case, in which the grid spacing is the same in all directions; that is, the \ipoint{number of points per direction} is given by $N_1=N_2=N_3=N$. Our actual implementation does not require this (see \secref{s:results} for an example). We approximate \[\st(x)=\sum_{\vect{k}} \hat{\st}_{\vect{k}}\exp(-\vect{k}\cdot\po),\] where $\vect{k}=(k_1,k_2,k_3)\in\mathbb{N}^3$ is a multi-index with $-N/2+1 \leq k_j \leq N/2, j =1,2,3$. The corresponding \ipoint{regular grid points} are given by $\po_{\vect{i}}=2\pi\vect{i}/N$, where $\vect{i}=(i_1,i_2,i_3)\in\mathbb{N}^3$ and $0\leq i_j \leq N-1, j =1,2,3$.

We refer to $\{\hat{\st}_{\vect{k}}\}$ as the \ipoint{spectral coefficients} of $\rho$. Mappings between $\{\st_{\vect{i}}\}$ and $\{\hat{\st}_{\vect{k}}\}$ are done using the forward and inverse \ipoint{Fast Fourier Transform} ({\bf FFT}). We use a similar spectral discretization for $\ad$ and for each component of the velocity field $\ve$. All derivatives are performed by first taking the FFT and then filtering the spectral coefficients appropriately. In general, the input images $\st_R$ and $\st_T$  may not be periodic functions. In that case a spectral approximation will create excessively high aliasing errors. To address this, we use zero-padding for $\st_R$ and $\st_T$. Also, in general, images will have discontinuities and thus are not differentiable, creating similar aliasing problems. So, before applying our algorithm, we smooth them spectrally with a Gaussian filter whose bandwidth is $2\pi/N$ (the grid size). Notice that our spectral representation with periodic boundary conditions allows us to apply all the different spatial operators---including $\Lap^{-1}$ and $\ivbi$---in a stable, accurate, and extremely efficient manner. As a result, the main cost of the computation will be solving the transport equations, not applying and inverting  elliptic differential operators.

\subsubsection{Time}\label{s:time}

We choose a \ipoint{semi-Lagrangian} method since it is unconditionally stable~\cite{Falcone:1998a} and allows us to take a small number of time steps $n_t\in\mathbb{N}$. This is critical since we store several space-time fields.  For example, when solving \eqref{e:inc-state} for $\tilde{\st}(t)$, we need $\st(t)$ for all $t$. For large $n_t$ the storage requirements become excessive and more sophisticated checkpointing schemes~\cite{Akcelik:2002a} are required---which are more expensive. If we were using a Courant-Friedrichs-Lewy ({\bf CFL}) restricted\footnote{The CFL condition defines an upper bound on the time step size to ensure a stable solution of stiff, time-dependent PDEs~\cite{LeVeque:1992a}.} scheme for solving the transport equations, storing the time history would have been impossible, since we had to store hundreds of time steps (see, e.g., \cite{Mang:2015a,Mang:2016a,Mang:2016b}).

To explain the semi-Lagrangian method we reintroduce the notational dependence in space. We consider the general transport equation for a scalar field $\nu(\po,t)$ (with a stationary velocity) \[\p_t \nu(\po,t) + \ve(\po)\cdot\Grad\nu(\po,t) = f(\, \nu(\po,t),\po)\,).\] For example, for~\eqref{e:inc-adj} we have $\nu=\tilde{\ad}$ and $f=\tilde{\ad}\Div\ve+\Div(\ad\tilde{\ve})$. Then, for each $\po$, to compute $\nu(\po,\dt)$ given $\nu(\po,0)$, we first compute a new point $\Po$, \ipoint{the semi-Lagrangian point}, using the scheme below:
\begin{equation}\label{e:sml-trajectory}
\begin{split}
\Po_*&=\po-\dt\ve(\po);\\
\Po &= \po - \frac{\dt}{2}\left(\,\ve(\po) +
\ve(\Po_*)\,\right),
\end{split}
\end{equation}
\noindent and then we set
\begin{equation}\label{e:sml-advection}
\begin{split} \nu_0(\Po) &= \nu(\Po,0);\\
f_0(\Po) &= f(\nu_0(\Po),\Po);\\
\nu_*(\po) &= \nu_0(\Po)+\dt f_0(\Po);\\
f_*(\po) &= f(\nu_*(\po),\po);\\ \nu(\po,\dt) &= \nu_0(\Po) + \frac{\dt}{2}(f_0(\Po)+f_*(\po)).
\end{split}
\end{equation}

\noindent This scheme is fully explicit and \emph{unconditionally stable}. Recall that $\po$ is a regular grid point and thus $\nu(\po,0)$ and $\ve(\po)$ are known. $\Po_*$ and $\Po$ are not regular grid points. Computing $\ve$ and $\nu$ at these off-grid locations requires multiple interpolations: three interpolations for $\ve(\Po_*)$, interpolations for the $f$ terms that depend on the semi-Lagrangian point $\Po$, and finally one interpolation for $\nu(\Po,0)$. If $f$ depends on derivatives of $\nu$, we first differentiate on the regular grid and then we interpolate the derivatives. The same scheme is used for the adjoint equations by changing the time variable from $t$ to $\tau=1-t$, so that $-\p_t \ad(t) = \p_\tau \ad(\tau)$ and $\ad(t=1)=\ad(\tau=0)$. Note that the interpolation cannot be done using a FFT, since the interpolation points can be spaced irregularly between grid points.

Cubic interpolation is typically preferred, compared to linear interpolation, because the interpolation errors will be accumulated throughout the time stepping without a time-step factor~\cite{Boyd:2000a}. We use a \ipoint{tricubic interpolation scheme}, which we discuss in~\secref{s:interp}.

\subsection{Parallel algorithms}\label{s:hpc}

The main computational kernels are the 3D FFTs to compute derivatives, elliptic operators and their inverses, the interpolation to off-grid points needed for the semi-Lagrangian time stepping, the Krylov solver (PCG) for the Hessian, and the Newton solver. The 3D FFT has well-known algorithmic complexity. The interpolation on semi-Lagrangian points is the most expensive parts of the computation, despite being local. As it turns out, about 60\% of the overall time for the image registration problem is spent on interpolation. The Krylov and Newton solvers are sequential across iterations whereas all the function, gradient, and Hessian evaluations are done using data parallelism. Below we give more details on the FFT, the interpolation, and how we put everything together.

\begin{figure}
\centering
\includegraphics[width=0.8\linewidth]
{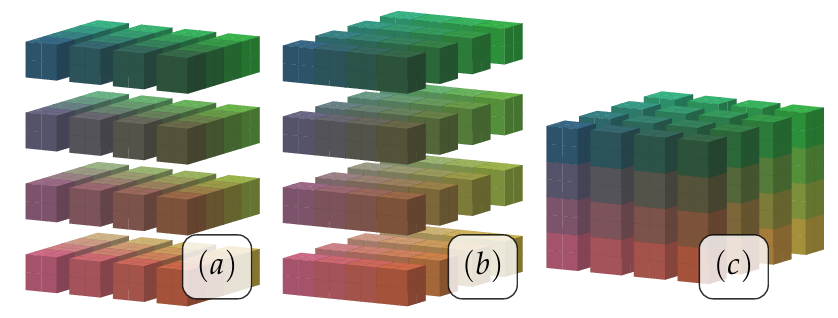}
\caption{Here we explain the data partitioning, which is based on the pencil decomposition for 3D FFTs~\cite{Karypis:2003a}. The colors indicate the data partitioning; each color corresponds to the data assigned to an MPI task. Subfigure (a) represents the input distribution of the (volumetric) image. After an FFT in the first coordinate, in (b) we do the FFT in the second coordinate. This requires $\sqrt{p}$ concurrent {\tt alltoall} between groups of $\sqrt{p}$ MPI tasks. This process is repeated for the third direction in (c) and has the same communication costs as (b) (image modified from~\cite{Gholami:2016a}).}
\label{f:pencil}
\end{figure}

\subsubsection{Partitioning and FFT}\label{s:partitioning}

Let $N_i$, $i=1,2,3$, be the number of grid points in the $i^{\rm th}$ dimension. Also assume we have $p = p_1 p_2$ MPI tasks. We partition the data using the \ipoint{pencil decomposition} of 3D FFT (see \figref{f:pencil}). Each MPI task gets $(N_1/p_1)\, (N_2/p_2)\, N_3$ grid points. There is no partitioning in time and all the time steps are stored in memory. The scalability of the 3D FFT has been well studied~\cite{Czechowski:2012a,Karypis:2003a}. The 3D FFT requires $\bigO(\frac{7.5N^3}{p}\log N)$ computations and $\bigO(\lat\sqrt{p}+\ban\frac{3N^3}{p})$ communications. We use the open-source package AccFFT~\cite{accfft-home-page,Gholami:2016b} that supports both GPU and CPU FFTs and is based on the 1D FFTs implemented in the FFTW package~\cite{fftw-home-page}. Our code features optimizations for the $\Grad$ and $\Div$ operators that allow us to avoid multiple 3D FFTs. For example, $\Grad \st = (\p_1 \st,\p_2 \st, \p_3 \st)$ requires $N_2 N_3$ 1D FFTs across the first coordinate, diagonal scaling, and then the same number of inverse FFTs again across the first dimension. A similar process is required for the other components but they require collective all-to-all communications for rearranging the data. The remaining operators $\lea$, $\vbi$, and $\ivbi$ are diagonal and require standard 3D FFTs.

\subsubsection{Interpolation}\label{s:interp}

For every grid point $\po_{\vect{i}}$ we have to find points $\Po_{\vect{i}}$ required in~\eqref{e:sml-advection} using~\eqref{e:sml-trajectory}. In the distributed case, every processor interpolates all the points that fall into the region defined by its pencil (that would be subfigure (a) in~\figref{f:pencil}). This is essentially an {\tt alltoallv} operation. We refer to this step as ``scatter'' phase. Note that the points need to be constructed only when the velocity field changes. In a Newton iteration for a given $\ve$ we have to compute these points only once for $\ve$ (forward transport) and for $-\ve$ (adjoint equations); that is the scatter phase needs to be done once per field per Newton iteration. This results in speedups due to savings in communication and computation. After the scatter phase is completed, each process has to perform a cubic interpolation on the points that it owns locally as well as the points it received from the other processors. After this step is done, an {\tt alltoallv} operation is necessary to send/recv all the interpolation results. This needs to be done once per time step.

The computation is organized as follows. For every forward or adjoint solve, we invoke an \ipoint{interpolation planner}, which performs the scatter phase and stores the semi-Lagrangian points and creates the communication plans for the transport equation. Then the actual transport~\eqref{e:sml-advection}, which involves multiple interpolations at every time step, is performed. For divergence-free $\ve$, the computation of~\eqref{e:varopt:state} and~\eqref{e:adjoint} involves only interpolations. For~\eqref{e:inc-state} and~\eqref{e:inc-adj} it also involves differential operators for the gradient and divergence operators that appear on the right-hand side.

Note that it is possible for an interpolation point to fall in-between the locally owned domains of the processes. This is because the local domain of each process is disjoint from others. For this reason, every processor maintains a layer of \ipoint{ghost points}, regular grid points that belong to other processors. The values of $\nu$ at these points must be synchronized before interpolation takes place. Notice that for every point we have to bring in 64 scalar values and perform roughly $10\times64$ floating point operations. The constant is related to the 64 coefficients ($4^3$) required to build and evaluate the tricubic interpolant times five flops per coefficient. Therefore, the computation to memory traffic ratio will be $\bigO(1)$---the computation is memory bound. Blocking, prefetching, and vectorization can be used to improve the performance.

\begin{algorithm}[H]
\caption{Parallel tricubic interpolation.}
\begin{algorithmic}[1]
	\renewcommand{\algorithmicrequire}{\textbf{Input:}}
	\renewcommand{\algorithmicensure}{\textbf{Output:}}
	\REQUIRE $\{\Pop\}\!\!\in\!\!\Omega_r$, {\tt owner}$(\Pop)$, $\{\poi\}\!\!\in\!\! \Omega_r$, {\tt worker}$(\poi)$, $\nu(\poi)$, MPI task $r$
	\ENSURE $\nu(\Poi)$
	\STATE Communicate $\nu$ values for ghost points.\label{a:mpi-interpolation:ghosts}
	\STATE Send/recv $\Pop$ from $r$ to/from {\tt owner}($\Pop$).\label{a:mpi-interpolation:send}
	\STATE Locally interpolate to compute $\nu(\Poi)$.\label{a:mpi-interpolation:interp}
	\STATE Send/recv $\nu(\Poi)$ to/from {\tt worker}($\poi$) to $r$.\label{a:mpi-interpolation:recv}
\end{algorithmic}
\label{a:mpi-interpolation}
\end{algorithm}

The execution flow of the interpolation algorithm is explained in \algref{a:mpi-interpolation}, for an MPI task $r$. Let $\nu(\poi)$ be the value of the scalar field $\nu$ at a regular grid point $\poi$. Also, let $\Pop$ be the corresponding semi-Lagrangian points for each $\ipr$ computed by~\eqref{e:sml-trajectory}. Let $\Omega_j$ be the spatial domain assigned to the MPI task $j$. As shown in figure~\ref{f:smlpoints}, $\Pop \in \Omega_r$ does not imply that $\po_{\ipr} \in \Omega_r$ and vice versa. For an off-grid point $\Pop$, {\tt owner}$(\Pop)$ computes the MPI task that owns that point. That is, $\Pop \in \Omega_r$ but $\po_{\ipr} \in \Omega_{{\mathtt owner}(\Pop)}$. Similarly, $\poi\in \Omega$ but $\Poi\in\Omega_{{\mathtt worker}(\Pop)}$. Of course for most points both the owner and worker domains will be identical to $\Omega_r$.  In line~\ref{a:mpi-interpolation:ghosts} every task sends values of $\nu$ that it owns to its four neighbors (recall that we use a pencil decomposition) with a communication cost of $4(\ban
N^2/p  + \lat))$ (the four corner neighbors can be combined with the messages of the edge neighbors, but appropriate ordering of the messages). In line~\ref{a:mpi-interpolation:send}, for $\Pop$ whose corresponding regular grid point $\poi$ belongs to a different MPI task, $r$ sends $\nu(\Pop)$ to that task.

In line~\ref{a:mpi-interpolation:interp} the interpolation takes place. This phase requires roughly $\bigO(600 N^3/p)$ floating point operations. This is followed by an {\tt alltoall} communication in line~\ref{a:mpi-interpolation:recv} to send/recv interpolation results. Our GPU implementation follows a similar strategy.

\subsubsection{Algorithm for incremental state equation}

We summarize the steps needed to solve the incremental forward problem~\eqref{e:inc-state} for one time step in algorithm~\ref{a:ts} to illustrate the communication and computation pattern.

\begin{algorithm}[H]
\caption{One time step of the solution of the incremental forward problem.}
\begin{algorithmic}[1]
    \renewcommand{\algorithmicrequire}{\textbf{Input:}}
    \renewcommand{\algorithmicensure}{\textbf{Output:}}
    \REQUIRE\! $\ve(\poi),\tilde{\ve}(\poi),\st(\poi,0),\st(\poi,\dt),\tilde{\st}(\poi,0),\Poi$
    \ENSURE\! $\tilde{\st}(\poi,\dt)$
    \STATE $\st_0(\Poi)=\st(\Poi,0)$ using \algref{a:mpi-interpolation}.\label{a:ts:itrp}
    \STATE $\Grad\st(\poi,0)$ using FFT.\label{a:ts:vFFT1}
    \STATE $f_0(\poi) = -\tilde{\ve}(\poi) \cdot \Grad\st(\poi,0)$.
    \STATE $f_0(\Poi)$ using $f_0(\poi)$ and \algref{a:mpi-interpolation}.\label{a:ts:vitrp}
    \STATE $\st_*(\poi) = \st_0(\Poi) + \dt f_0(\Poi)$.
    \STATE $\Grad \st_*(\poi)$ using FFT.\label{a:ts:vFFT2}
    \STATE $f_*(\poi)= -\tilde{\ve}(\poi)\cdot\Grad \st_*(\poi)$.
    \STATE $\tilde{\st}(\poi,\dt) = \st_0(\Poi) + \frac{\dt}{2}(f_0(\Poi) + f_*(\poi))$.
\end{algorithmic}
\label{a:ts}
\end{algorithm}

This calculation requires four interpolation steps using~\algref{a:mpi-interpolation}, one for the scalar interpolation in line~\ref{a:ts:itrp} and three for the vector interpolation in line~\ref{a:ts:vitrp}. It also requires four FFTs: two for line~\ref{a:ts:vFFT1} (it is two because we need to go the spectral domain, differentiate, and then back to the spatial domain) and two for line~\ref{a:ts:vFFT2}. The other parts are triple ``for-loops'' over all the grid points $\vect{i}$ in $\Omega_r$. The FFTs require global synchronizations. The \ipoint{total cost of the incremental adjoint solve} is four 3D FFTs and two interpolations. The incremental adjoint requires the same computations (for divergence-free velocity fields).

\subsubsection{Complexity of Hessian matvec and overall algorithm}

Every \ipoint{Hessian matvec} requires $n_t$ forward and adjoint solves or $8n_t$ FFTs and $4n_t$ interpolations. The remaining operations of applying the regularization and the preconditioner are negligible since they include just 2 FFTs each. The gradient is also cheaper since~\eqref{e:varopt:state} and~\eqref{e:adjoint} are simpler than the ones in~\eqref{e:hessian}. Regarding \ipoint{memory}, every task needs to store $2 n_t N^3/p + 5N^3/p$ values for the incremental adjoint and state variables. Therefore, accounting the complexities for the FFT and interpolation we obtain,
\begin{align*}
T_{\mathrm{flop}} \approx \;& n_t\left(8\frac{7.5N^3}{p}\log N + 4\frac{600 N^3}{p}\right) \\
T_{\mathrm{mpi}} \approx \;& 8n_t \left(3\lat\sqrt{p}+\ban\frac{3N^3}{p}\right) + 4n_t\left(\lat + \ban\frac{N^2}{p}\right)
\end{align*}

This estimate assumes that the semi-Lagrangian points are uniformly distributed across processors, however, this is not guaranteed and depends on the velocity field and the CFL number. In practice the interpolation is the predominant cost of the calculation, at least for the problem sizes we have tested. For fixed $\beta$ the number of Newton iterations are independent of the mesh size, the inversion of highly ill-conditioned operators is done in linear time.

\section{Results}\label{s:results}

\subsection{Experimental setup}

In this section, we give details on the experimental setup we used to test our solver.

\subsubsection{Images}

\begin{figure}
\centering
\includegraphics[width=0.95\columnwidth]
{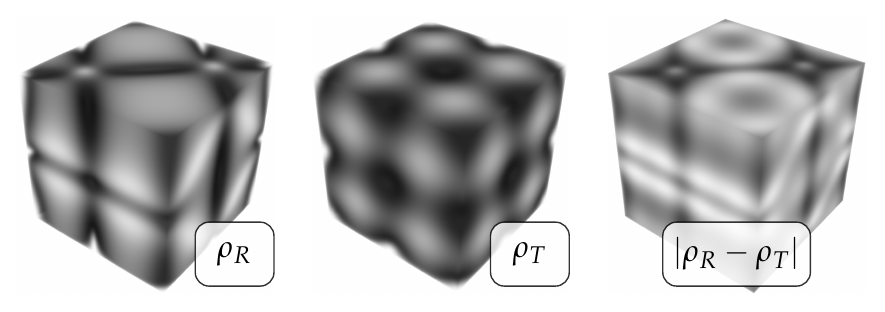}
\caption{3D visualization of a synthetic registration problem (volume rendering). From left to right: ($i$) reference image $\st_R$, ($ii$) template image $\st_T$, and ($iii$) initial (before registration) residual differences between $\st_R$ and $\st_T$. The reference image $\st_R$ is generated from $\st_T$ by solving the forward problem with a known velocity $\ve^\star$ (details can be found in the text). Dark areas indicate large residual differences and white areas zero residual differences.}
\label{f:smoothsynregprob}
\end{figure}

\begin{figure*}
\centering
\includegraphics[width=0.8\textwidth]
{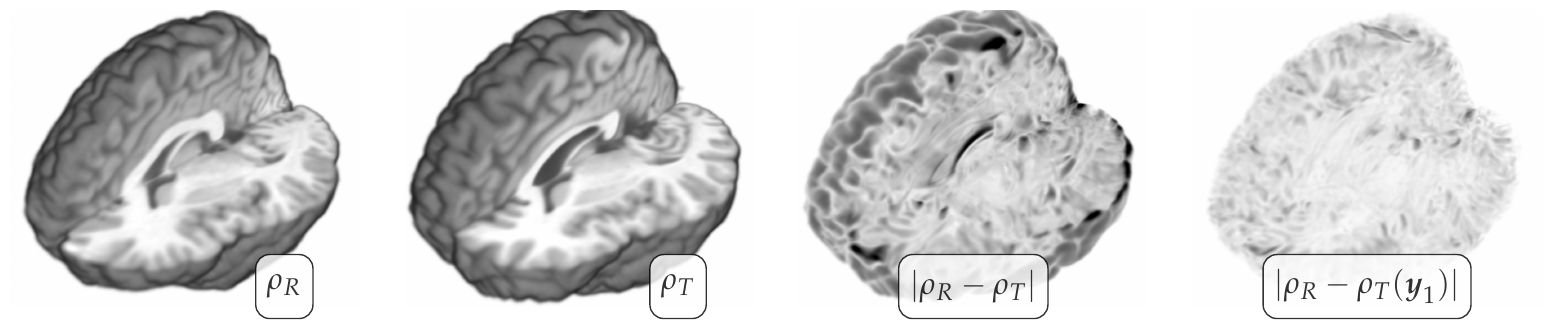}
\caption{3D visualization of the registration problem for the brain images. From left to right: ($i$) reference image $\rho_R$, ($ii$) template image $\rho_T$, ($iii$) the residual differences between $\rho_R$ and $\rho_T$ (before registration), and ($iv$) the residual differences between $\rho_R$ and $\rho_T(\vect{y}_1)$ (deformed template image; after registration). Dark areas indicate a large residual and white areas no residual differences.}
\label{f:brain3d}
\end{figure*}

We use one real-world and one synthetic image to test our algorithm. For the synthetic case we construct the template image as follows: $\st_T(\po) = (\sin^2(x_1) + \sin^2(x_2) + \sin^2(x_3))/3$; the velocity is given by $\ve^\star(\vect{x})=(\cos(x_1)\sin(x_2),\cos(x_2)\sin(x_1),\cos(x_1)\sin(x_3))^{\mathsf{T}}$; the reference image $\rho_R$ is the solution of~\eqref{e:varopt:state} with the exact velocity $\ve^\star$ (see \figref{f:smoothsynregprob} for an illustration of this problem).\footnote{For the incompressible case we use a similar but divergence free velocity field $\vect{v}^\star$.} We use a synthetic case to perform the scaling studies, since medical images come with a fixed resolution/grid size. To test our scheme on real medical images, we use two 3D MRI brain images of different individuals (``multi-subject registration problem''; grid size: $256\times300\times256$). This data is from the \emph{Non-rigid Registration Evaluation Project} ({\bf NIREP}) \cite{Christensen:2006a}.\footnote{The data is available at \href{http://nirep.org}{\tt http://nirep.org}; the interested reader is referred to~\cite{Christensen:2006a} for more details. We consider the first two datasets {\tt na01} and {\tt na02} from this repository.} (see \figref{f:brain3d} for an illustration).

\subsubsection{Implementation and Hardware}

Our code is implemented in {\tt C++} and uses MPI and the OpenMP library for multithreading.  The code is compiled with the Intel {\tt C++} compiler using the {\tt -O3} flag. Although we have GPU implementations both for the FFT and the interpolation, we have not used accelerators in the results we report here.  We carry out runtime experiments on the TACC's ``Maverick'' system. Each compute node contains dual, ten-core Intel Xeon  E5-2680 v2 (Ivy Bridge) processors running at 2.8GHz with 12.8GB/core of memory. Each node also has an NVIDIA Tesla K40 GPU accelerator. We also report large-scale runs on TACC's ``Stampede'' system (two eight-core Xeon E5-2680 v1 (Sandy Bridge) processors with 32GB host memory per node). As mentioned before, we use PETSc's TAO for the nonlinear optimization, vector operations of PETSc for vector linear operations, and AccFFT for the Fourier transforms. The basic interface to TAO is the functional, gradient, Hessian matvec, and preconditioner, as well as routines to select the tolerances for the nonlinear solver and for the Newton steps.

\subsubsection{Parameters}

The regularization parameter $\beta$ is set to \num{1E-2} for the scalability runs (for both, the synthetic and the real-world brain example). The number of time steps $n_t$ controls the accuracy and should be related to the CFL number. For simplicity and to be able to compare different cases, we have kept it fixed to $n_t=4$. The gradient tolerance is $\vect{g}_{\text{tol}}=\num{1E-2}$ unless otherwise stated. We use an inexact Newton method with quadratic forcing. We do not report continuation results; all the runs are done for a single (experimentally determined) value of $\beta$. We report ($i$) wall-clock times, ($ii$) communication times, as well as ($iii$) the time to solution for our method with respect to different registration problems and parameter settings. Since the problem is non-convex and we are not interested in high-accuracy solutions, we opt for a Gauss-Newton approximation.\footnote{Many image registration codes don't even compute gradients and the termination criterion is the number of iterations. Given the limitations in the resolution and the image quality a relative reduction of the gradient by 1\% is typically considered quite excessive.}

\subsection{Scalability using synthetic images}

\begin{table*}
\caption{Computational performance of our solver for the synthetic registration problem illustrated in~\figref{f:smoothsynregprob} on TACC's ``Maverick'' computing system. We neglect the incompressibility constraint for these runs. We report the time to solution, and the communication and execution times for the FFT and the interpolation, respectively (in seconds). We report timings as a function of the number of unknowns (in space), and the number of nodes and tasks. We use 16 tasks per node.}
\centering
\begin{tabular}{crrrLlllL}
\toprule
\multicolumn{5}{l}{}                                  & \multicolumn{2}{l}{FFT}                 & \multicolumn{2}{l}{interpolation}\\
\midrule
       & $N^3$   & nodes & tasks & time to solution   & communication      & execution          & communication      & execution\\
\midrule
\runid & $64^3$  &    1  &    16 & \num{1.537362e+00} & \num{1.197925e-01} & \num{9.690595e-02} & \num{1.822541e-01} & \num{8.200352e-01}\\
\runid &         &    2  &    32 & \num{9.500799e-01} & \num{1.421411e-01} & \num{4.884005e-02} & \num{1.149702e-01} & \num{4.265692e-01}\\
\midrule
\runid & $128^3$ &    1  &    16 & \num{1.517810e+01} & \num{1.733428e+00} & \num{1.346874e+00} & \num{1.835548e+00} & \num{6.656434e+00}\\
\runid &         &    2  &    32 & \num{7.878994e+00} & \num{1.296599e+00} & \num{5.466156e-01} & \num{1.167361e+00} & \num{3.487624e+00}\\
\runid &         &    4  &    64 & \num{4.701970e+00} & \num{1.185951e+00} & \num{2.834654e-01} & \num{5.434339e-01} & \num{1.871287e+00}\\
\runid &         &   16  &   256 & \num{2.006127e+00} & \num{6.680493e-01} & \num{6.601548e-02} & \num{1.860487e-01} & \num{4.908004e-01}\\
\midrule
\runid & $256^3$ &    2  &    32 & \num{7.988984e+01} & \num{1.443580e+01} & \num{1.005647e+01} & \num{1.079196e+01} & \num{2.831862e+01}\\
\runid &         &    8  &   128 & \num{2.303933e+01} & \num{7.269954e+00} & \num{1.562002e+00} & \num{2.597237e+00} & \num{8.038670e+00}\\
\runid &         &   32  &   512 & \num{7.232803e+00} & \num{2.673439e+00} & \num{3.375719e-01} & \num{5.928009e-01} & \num{2.000637e+00}\\
\runid &         &   64  &  1024 & \num{4.718962e+00} & \num{1.700525e+00} & \num{1.722865e-01} & \num{4.798732e-01} & \num{1.035110e+00}\\
\midrule
\runid & $512^3$ &    8  &   128 & \num{1.914122e+02} & \num{4.495866e+01} & \num{2.375949e+01} & \num{2.175900e+01} & \num{6.891822e+01}\\
\runid &         &   32  &   512 & \num{6.072784e+01} & \num{1.897447e+01} & \num{4.176254e+00} & \num{4.218898e+00} & \num{1.742407e+01}\\
\runid &         &   64  &  1024 & \num{3.289293e+01} & \num{1.281040e+01} & \num{1.765941e+00} & \num{2.326113e+00} & \num{8.571161e+00}\\
\bottomrule
\end{tabular}
\label{t:scalability-cpu-synthetic-maverick}
\end{table*}

\begin{table*}
\caption{Computational performance of our solver for the synthetic registration problem illustrated in~\figref{f:smoothsynregprob} on TACC's ``Stampede'' computing system. We neglect the incompressibility constraint for these runs. We report the time to solution, and the communication and execution times for the FFT and the interpolation, respectively (in seconds). We report timings as a function of the number of unknowns (in space), and the number of nodes and tasks. We use 2 tasks per node.}
\centering
\begin{tabular}{crrrLlllL}
\toprule
\multicolumn{5}{l}{}                                      & \multicolumn{2}{l}{FFT}                 & \multicolumn{2}{l}{interpolation}\\
\midrule
       & $N^3$    & nodes  & tasks   & time to solution   & communication      & execution          & communication      & execution\\
\midrule
\runid & $512^3$  & 256    & 512     & \num{3.835431e+01} & \num{4.605449e+00} & \num{2.624375e+00} & \num{4.116140e+00} & \num{1.979749e+01}\\
\runid &          & 512    & 1024    & \num{2.019381e+01} & \num{2.232177e+00} & \num{1.303958e+00} & \num{2.384872e+00} & \num{9.417384e+00}\\
\runid &          & 1024   & 2048    & \num{1.311402e+01} & \num{1.688897e+00} & \num{6.289797e-01} & \num{1.247142e+00} & \num{4.829060e+00}\\
\midrule
\runid & $1024^3$ & 256    & 512     & \num{3.535165e+02} & \num{3.289795e+01} & \num{3.103372e+01} & \num{3.724428e+01} & \num{1.927433e+02}\\
\runid &          & 512    & 1024    & \num{1.687821e+02} & \num{2.229121e+01} & \num{1.394521e+01} & \num{1.788936e+01} & \num{8.852654e+01}\\
\runid &          & 1024   & 2048    & \num{8.572117e+01} & \num{1.152305e+01} & \num{6.751315e+00} & \num{8.777678e+00} & \num{4.416057e+01}\\
\bottomrule
\end{tabular}
\label{t:scalability-cpu-synthetic-stampede}
\end{table*}

We use a fixed set of parameters, which we experimentally determined to yield a good balance between computational complexity and computational performance. We illustrate the registration problem in \figref{f:smoothsynregprob}. We report results for different grid resolutions ($N_i\in\{64,128,256,512,1024\}$), and different numbers of cores and MPI task configurations ($p\in\{1,2,4,8,16,64,256\}$). The results are reported in~\tabref{t:scalability-cpu-synthetic-maverick} (``Maverick'' runs) and \tabref{t:scalability-cpu-synthetic-stampede} (large scale ``Stampede'' runs). First, we interpret the $256^3$ runs (\#1--\#3), which represents a strong scaling analysis (in general, in image registration, strong scaling is what we're most interested in). From 32 tasks to 512 tasks the parallel efficiency is 67\%, whereas from 32 to 1024, the efficiency is 50\%. This is not ideal---however, it is quite good. The majority of the calculation for low task counts goes to the interpolation computation, whereas, as we increase the number of tasks, the majority of time goes to the FFT communication phase.\footnote{We use FFTs for the discretization of differential operators since this allows us to invert them at the cost of a spectral diagonal scaling. This offers the opportunity to exactly fulfill and \emph{efficiently} eliminate the incompressibility constraint from the optimality system. Also, it allows for an efficient preconditioning of the Hessian with essentially no construction cost (see \secref{s:methods} for details).} Similar conclusions can be drawn for the $128^3$ set; again, going from 16 tasks to 256 tasks, we observe 50\% efficiency. For the $512^3$ (\#11--\#13) the efficiency is 72\%. The latter is a problem with 1.5 billion unknowns for the velocity, without counting the unknowns for the state and adjoint fields; it only takes 32 seconds to solve to an accuracy of practical interest.

If we look the weak scaling results, we can consider runs \#3, \#8, and \#13, in which we increase the problem size by a factor of eight and the number of tasks also by a factor of eight. The overall timings are 15.2 seconds, 23 seconds, and 32 seconds, respectively, which again is not perfect. If we look more closely at how the time is allocated, we observe that the execution time for the FFT scales perfectly in these three runs (1.35 seconds, 1.56 seconds, and 1.77 seconds, respectively). The interpolation execution also scales well, both in terms of communication and computation. The deterioration of the overall time is due to the FFT communication costs. The largest problem we solved for the synthetic case was run \#19, in which we have 3.2 billion unknowns for the velocity field on 2048 MPI tasks on ``Stampede''. It took 85 seconds. The good scalability of the computation phase confirms the algorithmic optimality of the preconditioned Newton--Krylov method. We report results for the incompressible case in \tabref{t:scalability-cpu-synthetic-maverick-ic}.

\begin{table*}
\caption{Computational performance of our solver for a synthetic registration problem similar to the one illustrated in~\figref{f:smoothsynregprob} on TACC's ``Maverick'' computing system. We use the incompressibility constraint for these runs (mass preserving diffeomorphism). We report the time to solution, and the communication and execution times for the FFT and the interpolation, respectively (in seconds). We report results for a fixed grid size ($128^3$) as a function of the number of nodes and tasks. We use 2 tasks per node.}
\centering
\begin{tabular}{crrLlllL}
\toprule
\multicolumn{4}{c}{}                         & \multicolumn{2}{l}{FFT} & \multicolumn{2}{l}{interpolation}\\
\midrule
       & nodes & tasks  & time to solution   & communication           & execution          & communication      & execution\\
\midrule
\runid &     1 &     1  & \num{1.484652e+02} & \num{0}                 & \num{1.977852e+01} & \num{2.816136e+00} & \num{9.259727e+01}\\
\runid &     2 &     4  & \num{4.273114e+01} & \num{3.177971e+00}      & \num{5.734492e+00} & \num{8.391321e-01} & \num{2.314790e+01}\\
\runid &     4 &     8  & \num{2.250517e+01} & \num{2.172877e+00}      & \num{2.721070e+00} & \num{5.828967e-01} & \num{1.145883e+01}\\
\runid &     8 &    16  & \num{1.087921e+01} & \num{1.103788e+00}      & \num{1.245564e+00} & \num{4.033411e-01} & \num{5.797230e+00}\\
\runid &    16 &    32  & \num{5.693886e+00} & \num{6.690867e-01}      & \num{6.204312e-01} & \num{2.683859e-01} & \num{2.928745e+00}
\\\bottomrule
\end{tabular}
\label{t:scalability-cpu-synthetic-maverick-ic}
\end{table*}

\subsection{Real-world registration problem}

\begin{figure*}
\centering
\includegraphics[width=0.8\textwidth]
{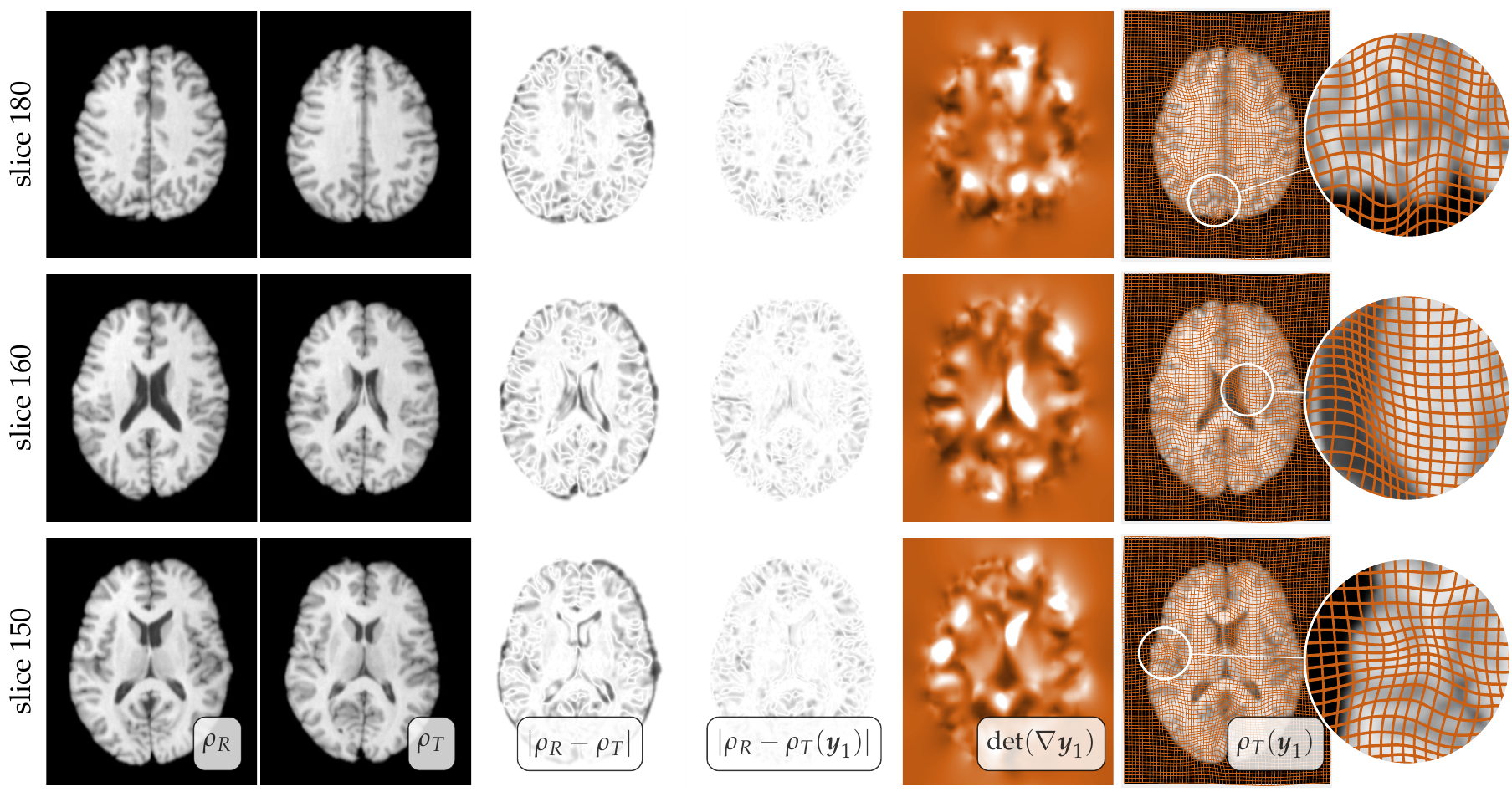}
\caption
{Exemplary registration results for the brain data sets. We display, from left to right axial slices of ($i$) the reference image $\rho_R$, ($ii$) the template image $\rho_T$, the residual differences ($iii$) between $\rho_R$ and $\rho_T$ before registration and ($iv$) after registration, ($v$) a point-wise map of the determinant of the deformation gradient (the color map represents volume change ranging from 0 to 2, where $\det(\Grad\df_1) = 0$ is black, $\det(\Grad\df_1)\in(0,2)$ corresponds to different shades of orange (from dark to bright), and $\det(\Grad\df_1)\geq 2$ is white), and ($vi$) the deformed template image $\rho_T(\df_1)$ with a grid in overlay (closeup to the right). The values for the determinant of the deformation gradient are strictly positive (i.e., the deformation map is diffeomorphic).}
\label{f:exemplary-nirep-results}
\end{figure*}

\begin{table*}
\caption{Strong scaling results for the brain images computed on ``Maverick''. We set the regularization parameter to $\beta=\num{1E-2}$. We perform two Newton iterations for these scalability runs. We report the number of nodes, the number of MPI tasks, and the communication and execution times for the FFT and the interpolation (in seconds).}
\centering
\begin{tabular}{crrLlllL}
\toprule
\multicolumn{4}{c}{}                         & \multicolumn{2}{l}{FFT} & \multicolumn{2}{l}{interpolation}\\
\midrule
       & nodes & tasks  & time to solution   & communication           & execution          & communication      & execution\\
\midrule
\runid &     1 &     1  & \num{1.336037e+03} & \num{0.000000e+01}      & \num{2.592920e+02} & \num{2.696116e+01} & \num{7.716794e+02}\\
\runid &     2 &     4  & \num{3.923728e+02} & \num{2.755865e+01}      & \num{6.909051e+01} & \num{5.730526e+00} & \num{1.903600e+02}\\
\runid &     8 &    16  & \num{9.544886e+01} & \num{8.585997e+00}      & \num{1.384455e+01} & \num{1.200038e+00} & \num{4.778420e+01}\\
\runid &    16 &    32  & \num{4.853950e+01} & \num{4.940710e+00}      & \num{6.499130e+00} & \num{5.353398e-01} & \num{2.364974e+01}\\
\runid &    32 &   256  & \num{1.203546e+01} & \num{4.025575e+00}      & \num{1.104590e+00} & \num{8.774614e-02} & \num{3.312572e+00}
\\\bottomrule
\end{tabular}
\label{t:strong-scaling-brain}
\end{table*}

\begin{table}
\caption{Sensitivity of the computational work load with respect to varying regularization weights $\beta\in\{\num{1E-2},\num{1E-3},\num{1E-4}\}$. We report results for four Newton iterations for the brain images. We report the number of Hessian matvecs and the time to solution (in seconds) and in parenthesis its relative increase from the base case.}
\centering

\begin{tabular}{ccRrl}
\toprule
         & $\beta$    & matvecs & time to solution~(~relative~) \\
\midrule
  \runid & \num{1E-1} & 43      & \num{2.423281e+01}~(~\z1.0~)  \\
  \runid & \num{1E-3} & 217     & \num{1.108109e+02}~(~\z4.6~)  \\
  \runid & \num{1E-5} & 1689    & \num{8.580886e+02}~(~35.0~)   \\
\bottomrule
\end{tabular}
\label{t:beta-dependence}
\end{table}

We report exemplary results for the brain data sets illustrated in \figref{f:brain3d} (grid size: $256\times300\times256$). We set the $\vect{g}_{\text{tol}}$ to \num{1E-2} and the maximal number of outer iterations (Newton steps) to 50, and $\beta=\num{1E-4}$. We study strong scaling and the sensitivity of the convergence of our solver with respect to changes in the regularization weight $\beta$. We report scalability results for the brain images in~\tabref{t:strong-scaling-brain}.  We display exemplary result for the considered datasets in \figref{f:exemplary-nirep-results}. We report results for varying choices of the regularization parameter $\beta$ in \tabref{t:beta-dependence}.

We observe that we can significantly reduce the computational timings if we switch to parallel architectures. The scaling results are consistent with what we observed for the synthetic data sets. We can reduce the wall clock time by two orders of magnitude if we change from one task on one node to 64 MPI tasks on 32 nodes. We can fit the entire problem on one node. This demonstrates the practicability of our solver. The communication and execution times of the FFT and the interpolator drop significantly as we increase the number of nodes. The interpolation time contributes again critically (about or more than 50\% of the time to solution).

As for the sensitivity with respect to the regularization parameter we can see that the number of Hessian matvecs (a proxy for the overall Newton-Krylov iterations) increases, as we reduce the regularization parameter from $\beta=\num{1E-3}$ to $\beta=\num{1E-5}$. The time to solution increases by a factor of 35 for the smallest $\beta$ reported here. This clearly demonstrates that the performance of our preconditioner is not ideal; it deteriorates with a reduction in $\beta$. As we have seen in the former section, the solver behaves independent of the mesh size. Implementing an improved scheme for preconditioning the Hessian requires more work.

\section{Conclusion}\label{s:conclusion}

We presented a complete algorithm for large deformation diffeomorphic medical image registration. We were able to solve problems of unprecedented scale. One may ask how such runs translate to a clinical setting. As the cost of computing drops, we are hopeful that 32- and 256-task calculations will be possible at a modest cost.

The proposed algorithm is flexible and scalable. It supports different types of regularization functionals and can be extended to different image distance measures. Our approach can be easily extended to vector images and---with some additional work---can also be extended to non-stationary (time-varying) velocities~\cite{Hart:2009a,Beg:2005a}. This will be necessary to register time-series of images or optical flow problems. All the parallelism related issues remain the same. A major remaining challenge is the design of preconditioners that are insensitive to the regularization parameter.

Finally, our algorithm relates to other applications besides medical imaging. For example applications in weather prediction and ocean physics (for tracking Lagrangian tracers in the oceans)~\cite{Kalnay:2002a}, for reconstruction of porous media flows~\cite{Fohring:2014a}, and registration of Micro-CTs for material science and biology~\cite{Ho:2006a}. Although our method is highly optimized for regular grids with periodic boundary conditions, many aspects of our algorithm carry over.


\begin{thebibliography}{10}
\providecommand{\url}[1]{#1}
\csname url@samestyle\endcsname
\providecommand{\newblock}{\relax}
\providecommand{\bibinfo}[2]{#2}
\providecommand{\BIBentrySTDinterwordspacing}{\spaceskip=0pt\relax}
\providecommand{\BIBentryALTinterwordstretchfactor}{4}
\providecommand{\BIBentryALTinterwordspacing}{\spaceskip=\fontdimen2\font plus
\BIBentryALTinterwordstretchfactor\fontdimen3\font minus
  \fontdimen4\font\relax}
\providecommand{\BIBforeignlanguage}[2]{{%
\expandafter\ifx\csname l@#1\endcsname\relax
\typeout{** WARNING: IEEEtranS.bst: No hyphenation pattern has been}%
\typeout{** loaded for the language `#1'. Using the pattern for}%
\typeout{** the default language instead.}%
\else
\language=\csname l@#1\endcsname
\fi
#2}}
\providecommand{\BIBdecl}{\relax}
\BIBdecl

\bibitem{Adavani:2008b}
S.~S. Adavani and G.~Biros, ``Fast algorithms for source identification
  problems with elliptic {PDE} constraints,'' \emph{SIAM Journal on Imaging
  Sciences}, vol.~3, no.~4, pp. 791--808, 2008.

\bibitem{Akcelik:2002a}
V.~Akcelik, G.~Biros, and O.~Ghattas, ``Parallel multiscale
  {G}auss-{N}ewton-{K}rylov methods for inverse wave propagation,'' in
  \emph{Proc ACM/IEEE Conference on Supercomputing}, 2002, pp. 1--15.

\bibitem{Amit:1994a}
Y.~Amit, ``A nonlinear variational problem for image matching,'' \emph{SIAM
  Journal on Scientific Computing}, vol.~15, no.~1, pp. 207--224, 1994.

\bibitem{Ashburner:2007a}
J.~Ashburner, ``A fast diffeomorphic image registration algorithm,''
  \emph{NeuroImage}, vol.~38, no.~1, pp. 95--113, 2007.

\bibitem{Ashburner:2011a}
J.~Ashburner and K.~J. Friston, ``Diffeomorphic registration using geodesic
  shooting and {G}auss-{N}ewton optimisation,'' \emph{NeuroImage}, vol.~55,
  no.~3, pp. 954--967, 2011.

\bibitem{Avants:2011a}
B.~B. Avants, N.~J. Tustison, G.~Song, P.~A. Cook \emph{et~al.}, ``A
  reproducible evaluation of {ANTs} similarity metric performance in brain
  image registration,'' \emph{NeuroImage}, vol.~54, pp. 2033--2044, 2011.

\bibitem{petsc-web-page}
\BIBentryALTinterwordspacing
S.~Balay, S.~Abhyankar, M.~F. Adams, J.~Brown \emph{et~al.}, ``{PETS}c {W}eb
  page.'' [Online]. Available: \url{http://www.mcs.anl.gov/petsc}
\BIBentrySTDinterwordspacing

\bibitem{Balay:2016a}
\BIBentryALTinterwordspacing
------, ``{PETS}c users manual,'' Argonne National Laboratory, Tech. Rep.
  ANL-95/11 - Revision 3.7, 2016. [Online]. Available:
  \url{http://www.mcs.anl.gov/petsc}
\BIBentrySTDinterwordspacing

\bibitem{Beg:2005a}
M.~F. Beg, M.~I. Miller, A.~Trouv\'e, and L.~Younes, ``Computing large
  deformation metric mappings via geodesic flows of diffeomorphisms,''
  \emph{International Journal of Computer Vision}, vol.~61, no.~2, pp.
  139--157, 2005.

\bibitem{Borzi:2012a}
A.~Borz\`{i} and V.~Schulz, \emph{Computational optimization of systems
  governed by partial differential equations}.\hskip 1em plus 0.5em minus
  0.4em\relax Philadelphia, Pennsylvania, US: SIAM, 2012.

\bibitem{Boyd:2000a}
J.~P. Boyd, \emph{Chebyshev and {F}ourier spectral methods}.\hskip 1em plus
  0.5em minus 0.4em\relax Mineola, New York, US: Dover, 2000.

\bibitem{Burger:2013a}
M.~Burger, J.~Modersitzki, and L.~Ruthotto, ``A hyperelastic regularization
  energy for image registration,'' \emph{SIAM Journal on Scientific Computing},
  vol.~35, no.~1, pp. B132--B148, 2013.

\bibitem{Chen:2011a}
K.~Chen and D.~A. Lorenz, ``Image sequence interpolation using optimal
  control,'' \emph{Journal of Mathematical Imaging and Vision}, vol.~41, pp.
  222--238, 2011.

\bibitem{Christensen:2006a}
G.~E. Christensen, X.~Geng, J.~G. Kuhl, J.~Bruss \emph{et~al.}, ``Introduction
  to the non-rigid image registration evaluation project,'' in \emph{Proc
  Biomedical Image Registration}, vol. LNCS 4057, 2006, pp. 128--135.

\bibitem{Crippa:2007a}
G.~Crippa, ``The flow associated to weakly differentiable vector fields,''
  Ph.D. dissertation, University of Zurich, 2007.

\bibitem{Czechowski:2012a}
K.~Czechowski, C.~Battaglino, C.~McClanahan, K.~Iyer \emph{et~al.}, ``On the
  communication complexity of 3{D} {FFT}s and its implications for exascale,''
  in \emph{Proc ACM/IEEE Conference on Supercomputing}, 2012, pp. 205--214.

\bibitem{Eisenstat:1996a}
S.~C. Eisentat and H.~F. Walker, ``Choosing the forcing terms in an inexact
  {N}ewton method,'' \emph{SIAM Journal on Scientific Computing}, vol.~17,
  no.~1, pp. 16--32, 1996.

\bibitem{Eklund:2013a}
A.~Eklund, P.~Dufort, D.~Forsberg, and S.~M. LaConte, ``Medical image
  processing on the {GPU}--past, present and future,'' \emph{Medical Image
  Analysis}, vol.~17, no.~8, pp. 1073--1094, 2013.

\bibitem{Falcone:1998a}
M.~Falcone and R.~Ferretti, ``Convergence analysis for a class of high-order
  semi-{L}agrangian advection schemes,'' \emph{SIAM Journal on Numerical
  Analysis}, vol.~35, no.~3, pp. 909--940, 1998.

\bibitem{Fohring:2014a}
J.~Fohring, E.~Haber, and L.~Ruthotto, ``Geophysical imaging of fluid flow in
  porous media,'' \emph{SIAM Journal on Scientific Computing}, vol.~36, no.~5,
  pp. S218--S236, 2014.

\bibitem{fftw-home-page}
\BIBentryALTinterwordspacing
M.~Frigo and S.~G. Johnson, ``{FFTW} home page.'' [Online]. Available:
  \url{http://www.fftw.org}
\BIBentrySTDinterwordspacing

\bibitem{Gholami:2016b}
A.~Gholami, J.~Hill, D.~Malhotra, and G.~Biros, ``{AccFFT}: {A} library for
  distributed-memory {FFT} on {CPU} and {GPU} architectures,'' \emph{arXiv
  e-prints}, 2016, in review (arXiv preprint:
  \url{http://arxiv.org/abs/1506.07933}).

\bibitem{Gholami:2016a}
A.~Gholami, A.~Mang, and G.~Biros, ``An inverse problem formulation for
  parameter estimation of a reaction-diffusion model of low grade gliomas,''
  \emph{Journal of Mathematical Biology}, vol.~72, no.~1, pp. 409--433, 2016.

\bibitem{accfft-home-page}
\BIBentryALTinterwordspacing
A.~Gholami and G.~Biros, ``{AccFFT} home page.'' [Online]. Available:
  \url{http://www.accfft.org}
\BIBentrySTDinterwordspacing

\bibitem{Karypis:2003a}
A.~Grama, A.~Gupta, G.~Karypis, and V.~Kumar, \emph{An Introduction to parallel
  computing: {D}esign and analysis of algorithms}, 2nd~ed.\hskip 1em plus 0.5em
  minus 0.4em\relax Addison Wesley, 2003.

\bibitem{Gunzburger:2003a}
M.~D. Gunzburger, \emph{Perspectives in flow control and optimization}.\hskip
  1em plus 0.5em minus 0.4em\relax Philadelphia, Pennsylvania, US: SIAM, 2003.

\bibitem{Gurtin:1981a}
M.~E. Gurtin, \emph{An introduction to continuum mechanics}, ser. Mathematics
  in Science and Engineering.\hskip 1em plus 0.5em minus 0.4em\relax Academic
  Press, 1981, vol. 158.

\bibitem{Ha:2010a}
L.~Ha, J.~Kr{\"u}ger, S.~Joshi, and T.~C. Silva, ``Multi-scale unbiased
  diffeomorphic atlas construction on multi-{GPUs},'' \emph{GPU Computing Gems
  Emerald Edition}, vol.~1, pp. 771--791, 2010.

\bibitem{Haber:2004a}
E.~Haber and J.~Modersitzki, ``Numerical methods for volume preserving image
  registration,'' \emph{Inverse Problems}, vol.~20, pp. 1621--1638, 2004.

\bibitem{Hart:2009a}
G.~L. Hart, C.~Zach, and M.~Niethammer, ``An optimal control approach for
  deformable registration,'' in \emph{Proc IEEE Conference on Computer Vision
  and Pattern Recognition}, 2009, pp. 9--16.

\bibitem{Hinze:2009a}
M.~Hinze, R.~Pinnau, M.~Ulbrich, and S.~Ulbrich, \emph{Optimization with {PDE}
  constraints}.\hskip 1em plus 0.5em minus 0.4em\relax Berlin, DE: Springer,
  2009.

\bibitem{Ho:2006a}
S.~T. Ho and W.~D. Hutmacher, ``A comparison of micro {CT} with other
  techniques used in the characterization of scaffolds,'' \emph{Biomaterials},
  vol.~27, no.~8, pp. 1362--1376, 2006.

\bibitem{Ino:2005a}
F.~Ino, K.~Ooyama, and K.~Hagihara, ``A data distributed parallel algorithm for
  nonrigid image registration,'' \emph{Parallel Computing}, vol.~31, no.~1, pp.
  19--43, 2005.

\bibitem{Ito:2008a}
K.~Ito and K.~Kunisch, \emph{Lagrange multiplier approach to variational
  problems and applications}.\hskip 1em plus 0.5em minus 0.4em\relax
  Philadelphia, Pennsylvania, US: SIAM, 2008, vol.~15.

\bibitem{Shackleford:2013a}
G.~S. James~Shackleford, Nagarajan~Kandasamy, \emph{High performance deformable
  image registration algorithms for manycore processors}.\hskip 1em plus 0.5em
  minus 0.4em\relax Morgan Kaufmann, 2013.

\bibitem{Kakinuma:2015a}
R.~Kakinuma, N.~Moriyama, Y.~Muramatsu, S.~Gomi \emph{et~al.},
  ``Ultra-high-resolution computed tomography of the lung: Image quality of a
  prototype scanner,'' \emph{PloS one}, vol.~10, no.~9, p. e0137165, 2015.

\bibitem{Kalnay:2002a}
E.~Kalany, \emph{Atmospheric modeling, data assimilation and
  predictability}.\hskip 1em plus 0.5em minus 0.4em\relax Oxford University
  Press, 2002.

\bibitem{Klein:2009a}
A.~Klein, J.~Andersson, B.~A. Ardekani, J.~Ashburner \emph{et~al.},
  ``Evaluation of 14 nonlinear deformation algorithms applied to human brain
  {MRI} registration,'' \emph{NeuroImage}, vol.~46, no.~3, pp. 786--802, 2009.

\bibitem{Klein:2010a}
S.~Klein, M.~Staring, K.~Murphy, M.~A. Viergever, and J.~P.~W. Pluim,
  ``{ELASTIX}: {A} tollbox for intensity-based medical image registration,''
  \emph{Medical Imaging, IEEE Transactions on}, vol.~29, no.~1, pp. 196--205,
  2010.

\bibitem{LeVeque:1992a}
R.~J. LeVeque, \emph{Numerical methods for conservation laws}.\hskip 1em plus
  0.5em minus 0.4em\relax Springer, 1992, vol. 132.

\bibitem{Lions:1972a}
J.-L. Lions, \emph{Some aspects of the optimal control of distributed parameter
  systems}.\hskip 1em plus 0.5em minus 0.4em\relax Philadelphia, Pennsylvania,
  US: SIAM, 1972.

\bibitem{Liu:2009a}
Y.~Liu, A.~Fedorov, R.~Kikinis, and N.~Chrisochoides, ``Real-time non-rigid
  registration of medical images on a cooperative parallel architecture,'' in
  \emph{Proc IEEE International Conference on Bioinformatics and Biomedicine},
  2009, pp. 401--404.

\bibitem{Lorenzi:2013b}
M.~Lorenzi, N.~Ayache, G.~B. Frisoni, and X.~Pennec, ``{LCC}-demons: a robust
  and accurate symmetric diffeomorphic registration algorithm,''
  \emph{NeuroImage}, vol.~81, pp. 470--483, 2013.

\bibitem{Lusebrink:2013a}
F.~L{\"u}sebrink, A.~Wollrab, and O.~Speck, ``Cortical thickness determination
  of the human brain using high resolution 3{T} and 7{T MRI} data,''
  \emph{Neuroimage}, vol.~70, pp. 122--131, 2013.

\bibitem{Mang:2015a}
A.~Mang and G.~Biros, ``An inexact {N}ewton--{K}rylov algorithm for constrained
  diffeomorphic image registration,'' \emph{SIAM Journal on Imaging Sciences},
  vol.~8, no.~2, pp. 1030--1069, 2015.

\bibitem{Mang:2016a}
------, ``Constrained {$H^1$}-regularization schemes for diffeomorphic image
  registration,'' \emph{SIAM Journal on Imaging Sciences}, 2016, to appear
  (arXiv preprint: \url{http://arxiv.org/abs/1503.00757}).

\bibitem{Mang:2016b}
------, ``A {S}emi-{L}agrangian two-level preconditioned {N}ewton--{K}rylov
  solver for constrained diffeomorphic image registration,'' \emph{arXiv
  e-prints}, 2016, in review (arXiv preprint:
  \url{http://arxiv.org/abs/1604.02153}).

\bibitem{Modat:2010a}
M.~Modat, G.~R. Ridgway, Z.~A. Taylor, M.~Lehmann \emph{et~al.}, ``Fast
  free-form deformation using graphics processing units,'' \emph{Computer
  Methods and Programs in Biomedicine}, vol.~98, no.~3, pp. 278--284, 2010.

\bibitem{Modersitzki:2004a}
J.~Modersitzki, \emph{Numerical methods for image registration}.\hskip 1em plus
  0.5em minus 0.4em\relax New York: Oxford University Press, 2004.

\bibitem{Modersitzki:2009a}
------, \emph{{FAIR}: Flexible algorithms for image registration}.\hskip 1em
  plus 0.5em minus 0.4em\relax Philadelphia, Pennsylvania, US: SIAM, 2009.

\bibitem{Nocedal:2006a}
J.~Nocedal and S.~J. Wright, \emph{Numerical Optimization}.\hskip 1em plus
  0.5em minus 0.4em\relax New York, New York, US: Springer, 2006.

\bibitem{Schackleford:2010a}
J.~A. Schackleford, N.~Kandasamy, and G.~C. Sharp, ``On developing {B}-spline
  registration algorithms for multi-core processors,'' \emph{Physics in
  Medicine and Biology}, vol.~55, no.~21, pp. 6329--6351, 2010.

\bibitem{Shams:2010a}
R.~Shams, P.~Sadeghi, R.~A. Kennedy, and R.~I. Hartley, ``A survey of medical
  image registration on multicore and the {GPU},'' \emph{Signal Processing
  Magazine, IEEE}, vol.~27, no.~2, pp. 50--60, 2010.

\bibitem{Shen:2003b}
D.~Shen and C.~Davatzikos, ``Very high-resolution morphometry using
  mass-preserving deformations and {HAMMER} elastic registration,''
  \emph{NeuroImage}, vol.~18, no.~1, pp. 28--41, 2003.

\bibitem{Sotiras:2013a}
A.~Sotiras, C.~Davatzikos, and N.~Paragios, ``Deformable medical image
  registration: {A} survey,'' \emph{Medical Imaging, IEEE Transactions on},
  vol.~32, no.~7, pp. 1153--1190, 2013.

\bibitem{Starosolski:2015a}
Z.~Starosolski, C.~A. Villamizar, D.~Rendon, M.~J. Paldino \emph{et~al.},
  ``Ultra high-resolution in vivo computed tomography imaging of mouse
  cerebrovasculature using a long circulating blood pool contrast agent,''
  \emph{Scientific Reports}, vol.~5, no. 10178, 2015.

\bibitem{Temam:1977a}
R.~Temam, \emph{{N}avier--{S}tokes equations: {T}heory and numerical
  analysis}.\hskip 1em plus 0.5em minus 0.4em\relax North-Holland Pub. Co.,
  1977.

\bibitem{tomer-e14}
R.~Tomer, L.~Ye, B.~Hsueh, and K.~Deisseroth, ``Advanced {CLARITY} for rapid
  and high-resolution imaging of intact tissues,'' \emph{Nature protocols},
  vol.~9, no.~7, pp. 1682--1697, 2014.

\bibitem{Rehman:2009a}
T.~ur~Rehman, E.~Haber, G.~Pryor, J.~Melonakos, and A.~Tannenbaum, ``3d
  nonrigid registration via optimal mass transport on the {GPU},''
  \emph{Medical Image Analysis}, vol.~13, no.~6, pp. 931--940, 2009.

\bibitem{Vercauteren:2008a}
T.~Vercauteren, X.~Pennec, A.~Perchant, and N.~Ayache, ``Symmetric log-domain
  diffeomorphic registration: {A} demons-based approach,'' in \emph{Proc
  Medical Image Computing and Computer-Assisted Intervention}, vol. LNCS 5241,
  no. 5241, 2008, pp. 754--761.

\bibitem{Vercauteren:2009a}
------, ``Diffeomorphic demons: {E}fficient non-parametric image
  registration,'' \emph{NeuroImage}, vol.~45, no.~1, pp. S61--S72, 2009.

\bibitem{Warfield:2000a}
S.~K. Warfield, M.~Ferrant, X.~Gallez, A.~Nabavi \emph{et~al.}, ``Real-time
  biomechanical simulation of volumetric brain deformation for image guided
  neurosurgery,'' in \emph{Proc ACM/IEEE Conference on Supercomputing}, 2000,
  pp. 23--23.

\bibitem{Yin:2009a}
Y.~Yin, E.~A. Hoffman, and C.-L. Lin, ``Mass preserving nonrigid registration
  of {CT} lung images using cubic {B}-spline,'' \emph{Medical Physics},
  vol.~36, no.~9, pp. 4213--4222, 2009.

\bibitem{Zhang:2015a}
M.-Q. Zhang, L.~Zhou, Q.-F. Deng, Y.-Y. Xie \emph{et~al.},
  ``Ultra-high-resolution {3D} digitalized imaging of the cerebral
  angioarchitecture in rats using synchrotron radiation,'' \emph{Scientific
  Reports}, vol.~5, no. 14982, 2015.

\end{thebibliography}
\end{document}